\documentclass[naturemag,twocolumn]{revtex4-1}
\usepackage{graphicx}
\usepackage{dcolumn}
\usepackage{bm}
\usepackage{color}
\usepackage{soul}
\usepackage[english]{babel}

\begin{document}

\newcommand{\phlam}{$^1$}
\newcommand{\kit}{$^2$}
\newcommand{\iosb}{$^{3}$}
\newcommand{\psinst}{$^4$}
\newcommand{\DLR}{$^{5}$}
\newcommand{\desy}{$^{6}$}
\title{From self-organization in relativistic electron bunches to coherent synchrotron light: observation using a photonic time-stretch digitizer}

 \author{Serge Bielawski\phlam$^{,*}$,
 Edmund Blomley\kit, Miriam Brosi\kit, Erik Br\"undermann\kit, Eva Burkard\phlam$^,$\iosb, Cl\'ement Evain\phlam, Stefan Funkner\kit,  Nicole Hiller\kit$^,$\psinst, Michael J. Nasse\kit, Gudrun Niehues\kit,  El\'eonore Roussel\phlam, Manuel Schedler\kit, Patrik Sch\"onfeldt\kit$^,$\DLR, Johannes L. Steinmann\kit, Christophe Szwaj\phlam, Sophie Walther\kit$^,$\desy, and Anke-Susanne M\"uller\kit
}
 \affiliation{
 \phlam Univ. Lille, CNRS, UMR 8523 - PhLAM - Physique des Lasers, Atomes et Mol\'ecules,  Centre
  d'\'Etude Recherches et Applications (CERLA), F-59000 Lille, France.\\
\kit Karlsruhe Institute of Technology (KIT), D-76131 Karlsruhe, Germany\\
\iosb Present address Fraunhofer Institute of Optronics, System Technologies and Image Exploitation (IOSB), D-76275 Ettlingen, Germany\\
 \psinst Present address Paul Scherrer Institute (PSI), 5232 Villigen, Switzerland.\\
\DLR DLR (Deutsches Zentrum f\"ur Luft und Raumfahrt) Institute of Networked Energy Systems, Carl-von-Ossietzky-Str. 15, D-26129 Oldenburg, Germany\\
\desy Present address DESY (Deutsches Elektronen-Synchrotron),  Notkestr. 85, D-22607 Hamburg, Germany}

\email[Corresponding author : ]{serge.bielawski@univ-lille.fr}
\date{\today}
\maketitle
{\bf In recent and future synchrotron radiation facilities, relativistic
  electron bunches with {increasingly} high charge density are
  needed for producing brilliant light at various wavelengths, from X-rays to
  terahertz. {In such conditions, interaction of electrons
    bunches with their own emitted electromagnetic fields} leads to
  instabilities and spontaneous formation of complex spatial
  structures. 
     Understanding these instabilities is therefore key in most electron accelerators. However, investigations suffer from the lack of non-destructive recording tools for electron bunch shapes. In storage rings, most studies thus focus on the resulting emitted
  radiation. Here, we
  present measurements of the electric field in the immediate vicinity
  of the electron bunch in a storage ring, over many turns. For recording the ultrafast electric field, we designed a photonic time-stretch analog-to-digital converter with terasamples/second acquisition rate. We could thus 
  observe the predicted link 
  between spontaneous pattern formation and
  giant bursts of coherent synchrotron radiation in a storage
  ring.}

\section*{introduction}
\begin{figure}[htbp]
\includegraphics[width=8cm]{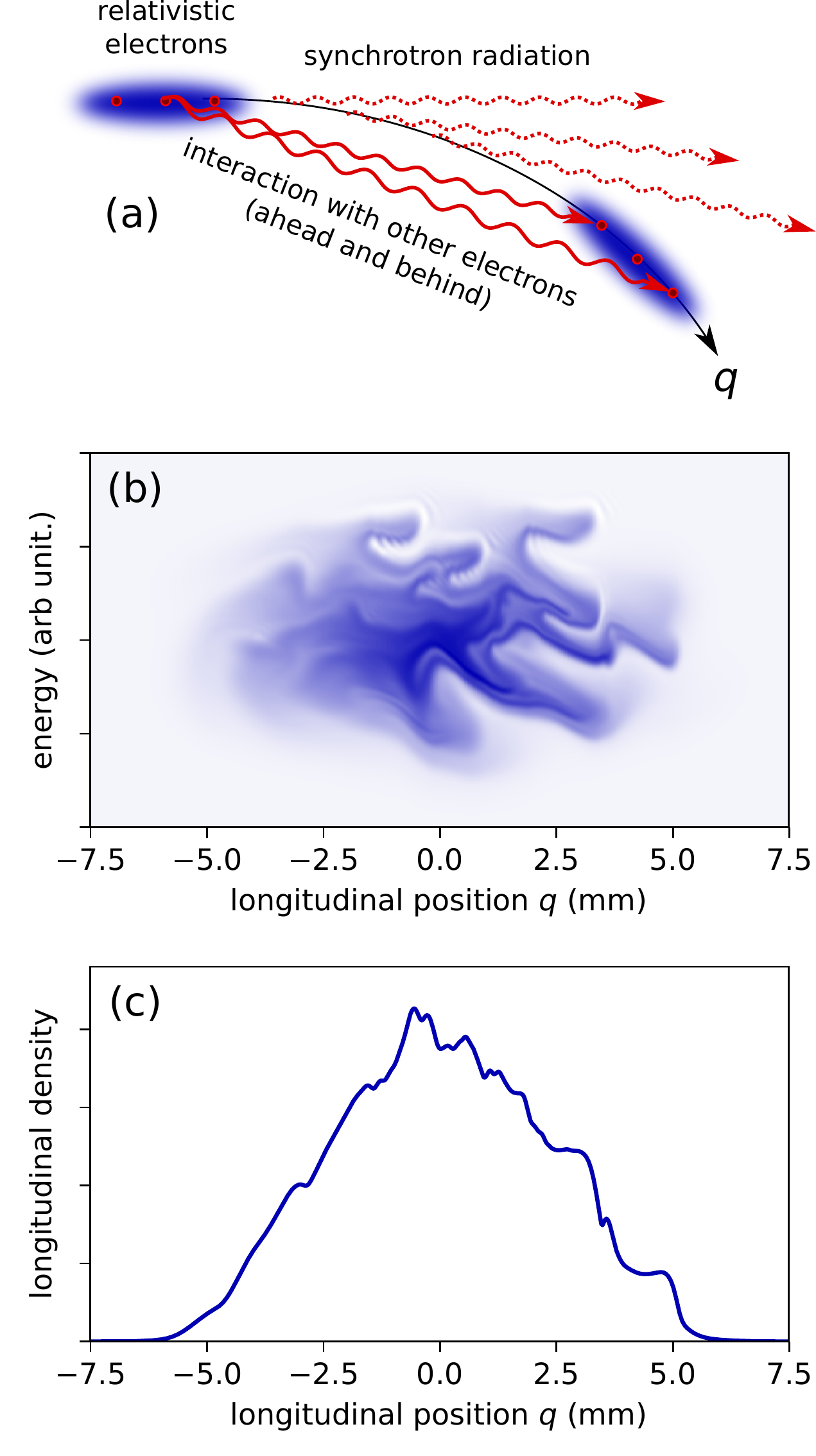}
\caption{{\bf Microbunching instability in a relativistic electron
  bunch.} {\bf (a)} Illustration: at accelerator locations where curved
  trajectories are present, each electron interacts with the coherent
  synchrotron radiation emitted by the others. {\bf (b,c)} Numerical
  simulation: resulting spontaneous appearance of a pattern in phase
  space, which evolves in a complex way (computation using KARA
  storage ring parameters, see Method Section, and Supplementary
  Video~1). {\bf (c)} corresponding longitudinal density profile. Note that
  the fast modulation in {\bf (c)} -- thought apparently small -- is responsible for a
  particularly intense emission of coherent synchrotron radiation
  (typically $10^3$--$10^5$ times the normal synchrotron radiation)
  ranging from the millimeter-wave to THz domains.}
\label{fig:fig0}
\end{figure}

Current storage ring synchrotron radiation facilities involve
challenges in photonics, both for understanding the light source and
for realizing suitable ultrafast measurement devices. Generation of
light for users is performed by using electron bunches in the
subnanosecond to picosecond range, with high charge density. {This density is so high that the light emitted by the electrons affects the dynamics} of neighboring electrons
in a dramatic way. 
    In particular, this nonlinear collective effect
leads to spontaneous formation of small-scale structures (in the
sub-millimeter to centimeter range) in the longitudinal profile of
electron bunches~\cite{hight1997ar,andersson2000coherent,byrd2002.PhysRevLett.89.224801,abo2002.PhysRevLett.88.254801,
takashima2005observation,katoh2007coherent,Karantzoulis2010300,muller2011metrology,Feikes2011metrology,cinque2011far, shields2012.1742-6596-357-1-012037,evain2012.0295-5075-98-4-40006,mullerexperimentalapects,barros2015characteristics,roy.RSI.84.033102.2013,PhysRevLett.114.204801,steinmann2018continuous}. This is known as the {\it microbunching instability}~\cite{venturini2002.PhysRevLett.89.224802,
    stupakov2002.PhysRevSTAB.5.054402,byrd2002.PhysRevLett.89.224801,abo2002.PhysRevLett.88.254801} (see Figure~\ref{fig:fig0})~. This effect is conceptually close to the
universal mechanisms of pattern formation in
Nature~\cite{cross1993.RevModPhys.65.851} due to interaction between
parts of the same system, {such as} the modulation instability in optical
fibers~\cite{tai1986observation,tai1986observation}, sand ripple
formation induced by the wind or under the
sea~\cite{charru2013sand}, 
or phantom traffic jams~\cite{helbing2001traffic}.

However, besides being a fascinating phenomenon of light and matter
self-organization, latest generation light sources must systematically
consider these collective effects for very practical
reasons. Spontaneous formation of small-scale microstructures can have
a deleterious effect on electron bunch stability and emission properties, and
they are at the same time a tremendous source of {coherent radiation in
the terahertz domain~\cite{byrd2002.PhysRevLett.89.224801,abo2002.PhysRevLett.88.254801,
takashima2005observation,katoh2007coherent,Karantzoulis2010300,muller2011metrology,Feikes2011metrology,cinque2011far, shields2012.1742-6596-357-1-012037,evain2012.0295-5075-98-4-40006,mullerexperimentalapects,barros2015characteristics,roy.RSI.84.033102.2013,PhysRevLett.114.204801,steinmann2018continuous}}, provided the instability can be mastered. This is the
reason why understanding and controlling the interplay between Coherent
Synchrotron Radiation (CSR) and the microbunching instability has
nowadays become a central open question in the development of
synchrotron radiation facilities.

To answer this question, it is essential to develop ultrafast photonic devices for electron bunch shape characterization. The challenges for the photonics community is high, given the need for
ultrashort (picosecond or femtosecond) temporal resolution, single-shot
operation, at high repetition rates (MHz and more), and given the particularly
challenging environment near relativistic electron
bunches. Recent advances consequently pushed 
photonics systems beyond the state of the art. {Ultrafast electric-field measurement techniques using femtosecond laser pulses (electro-optic sampling~\cite{EOS_first_spectral_encoding}) have allowed single-shot bunch shape measurement~\cite{PhysRevLett.88.124801}, and these techniques have then been extensively investigated and improved this last decade~\cite{casalbuoni2008numerical, steffen2009electro, schmidhammer2009single,PhysRevSTAB.15.070701,szwaj2016high}. As these techniques require compact femtosecond lasers, this also motivated a specific work on fiber-based sources, using parabolic pulse amplification~\cite{tamura1996pulse,fermann2000self}. This even led to new record spectral widths for parabolic pulse amplifiers~\cite{feurer2009ytterbium}}. 


{Ultrafast diagnostics also recently started to
use strategies from the emerging field of "photonic hardware accelerators"~\cite{jalali2015tailoring}, which aims at increasing the speed of electronic devices by combining them with specially designed photonic front-ends. In particular {\it photonic
  time-stretch analog to digital converters}~\cite{mahjoubfar2017time,time_stretch_first_bhushan1998time}
opened the way to the realization of ``single-shot terahertz oscilloscopes''~\cite{mahjoubfar2017time,roussel2015.EOS,kobayashi2016high,szwaj2016high,evain2017direct} providing up to tens of million traces per second.

The availability of new ultrafast measurement systems led to several
milestones in these storage ring investigations. Pioneer experiments
using a streak camera could visualize microstructures in the
several~GHz range at the VUV ring~\cite{kramermicrowave}. More
recently, electron bunch shapes have been indirectly characterized in
single-shot by using new detectors based on thin films of
superconducting YBCO~\cite{uvsor_MBI_ybco}, and high repetition rate
electro-optic sampling, using photonic
time-stretch~\cite{roussel2015.EOS}. Although this progress enabled to
record structures in single-shot up to the THz range, the obtained
information concerned only the far-field (i.e., the synchrotron
radiation) emitted by the microstructures~\cite{kramermicrowave,roussel2015.EOS,evain2017direct}.

In this article, we present a photonic system that enables to observe
microstructures and their evolution in a direct way, by monitoring the
electric field in the immediate vicinity of the electrons.

\section*{Results}
\subsection*{Experimental strategy}
Recording bunch shapes in a non-destructive way required two open problems to be solved. The first one
consisted in probing the electric field by approaching an electro-optic
crystal at few millimeters from the relativistic electron bunch
(Figure~\ref{fig:fig1}), without losing the electron bunch or damaging the
crystal. 
%
By carefully designing the experimental
setup~\cite{ANKA.EOS.IPAC13,thesis_hiller}, we could demonstrate the
possibility to operate the synchrotron facility with an electro-optic crystal
at 2-18 millimeters from the electron bunch. This pioneer experiment at the
ANKA (now KArlsruhe Research Accelerator -- KARA) storage ring thus opened the
way to real-time investigations of storage ring electron bunch shapes, under
the condition that a suitable photonic ultrafast readout system can be
designed. \begin{figure*}[htb]
\includegraphics[width=18.5cm]{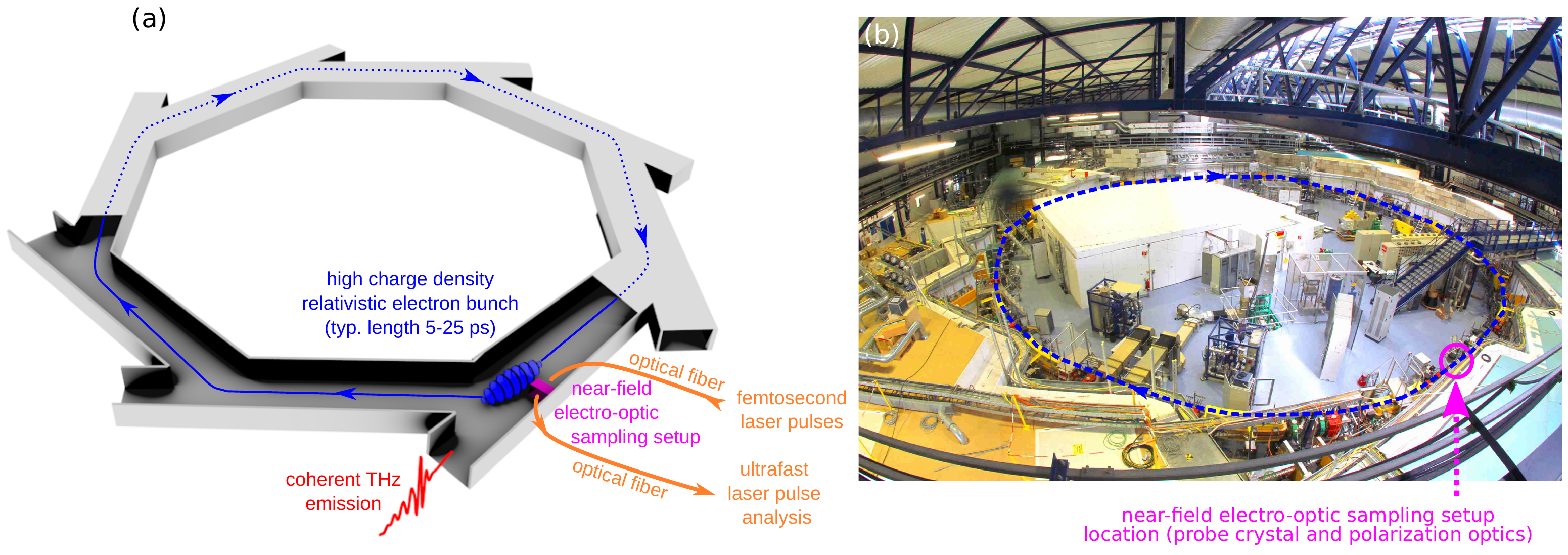} 
\caption{{\bf Global
strategy of the experiment (a), and picture of the KARA   storage ring (b).}
Interaction of a relativistic electron bunch with its own   emitted coherent
radiation leads to the so-called microbunching instability,   and formation of
a pattern with few millimeter period in the longitudinal   direction. For monitoring the
longitudinal electron bunch shape, we record the electric field evolution
in its vicinity (at few millimeters), using a specially designed
picosecond-speed {\it photonic-time-stretch analog-to-digital converter}. 
The digitization is made in two steps: (i) laser pulses are modulated by the electric field using an electro-optic crystal, and (ii) the modulated pulses are analyzed in single-shot, picosecond resolution, and multi-MHz acquisition rate. {Note that the crystal is actually placed above the electron beam (the whole photonic time-stretch digitizer is detailed in
Figure~\ref{fig:setup_eos})}. 
The electron bunch microstructure is also emitting intense coherent synchrotron radiation (CSR), which is simultaneously recorded.}
\label{fig:fig1} \end{figure*}

Optical readout of the crystal birefringence versus time was the second
key problem as measurements had to be be performed: (i) in single-shot, (ii)
with picosecond or sub-picosecond resolution, (iii) at several MHz acquisition rate. Moreover
this ultrafast readout needs to be performed with an important dynamical range because the
fast microstructures are expected to appear as a small modulation superimposed
on a large slowly-varying background (see
Figure~\ref{fig:fig0}c). At KARA, we have been exploring two
  directions in parallel. We have been developing a new generation of fast
  linear cameras (KALYPSO) with multi-MHz acquisition rates~\cite{rota2018kalypso,kalypso_microstructures_first,kehrer2018synchronous,funkner2018high}. In parallel, we have
  been developing a second direction consisting in using the so-called
  photonic time-stretch
  strategy~\cite{time_stretch_first_bhushan1998time,mahjoubfar2017time}. The latter
  strategy allows up to tens of MHz acquisition rate, using an association of
  commercial detectors and electronics. The results presented in this article
  are obtained with this strategy.

\subsection*{Photonic time-stretch Analog-to-Digital Converter}

{The photonic time-stretch digitizer setup is represented in Figure~\ref{fig:setup_eos}. The optical front-end combines two parts.} A
single-shot electro-optic sampling (EOS) system~\cite{ANKA.EOS.IPAC13}
imprints the electric field shape onto a chirped laser pulse~\cite{EOS_first_spectral_encoding,PhysRevLett.88.124801}. Then,
the laser pulse exiting the EOS system is stretched in a 2~km dispersive
fiber, so that the picosecond information is temporally stretched to
the nanosecond range, and can be recorded using a photodetector and a
conventional oscilloscope (5~GHz bandwidth is typically used here, see Methods). If we
start from a compressed laser pulse, the output signal should be a
replica of the ultrafast electric pulse, slowed-down by a factor~\cite{time_stretch_first_bhushan1998time,mahjoubfar2017time}:
\begin{equation}
M=1+\frac{L_2}{L_1},
\end{equation}
with $L_1$ and $L_2$ the lengths of fiber before and after the crystal
(if the fibers are identical). {Since an unknown
  amount of extra-dispersion is also present before the fiber of
  length $L_1$, we also measured the stretch factor experimentally. We
  found $M=75.8$, i.e., 1~nanosecond on the oscilloscope corresponds
  to a real duration of $13.2~$ps at the input for all results
  presented hereafter.}

\begin{figure*}[htbp]
\includegraphics[width=17cm]{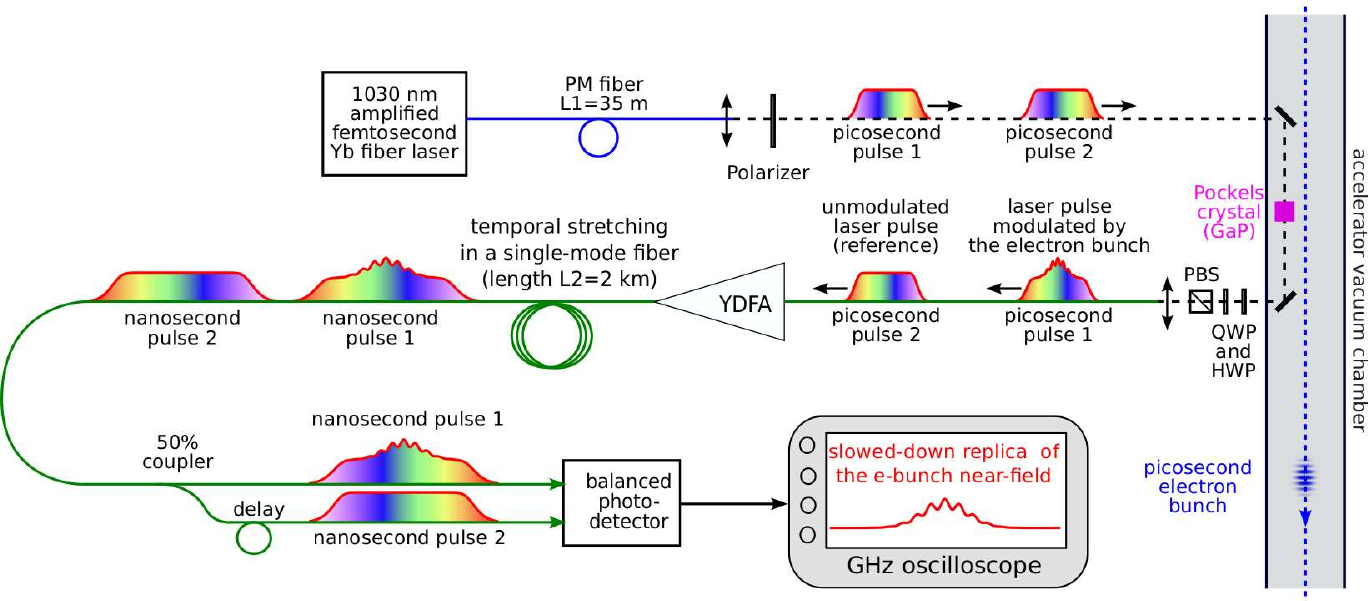} \caption{{\bf
Photonic time-stretch analog-to-digital converter realized for recording the
shape the of electron bunches at high repetition rate.} The electron bunch
near-field is imprinted onto a chirped laser pulse, by using   the Pockels
effect in a gallium phosphide (GaP) crystal. The laser pulse is then further
chirped in a long fiber, so that the modulation is slowed down to the
nanosecond range, and can be recorded by an oscilloscope. Furthermore, an
additional laser pulse which has not interacted with the electron bunch is
used as "zero field" reference, and is subtracted from the signal by a
balanced photodetector. Note that another reference laser pulse (not shown) is
also recorded and used in the offline data processing (see Methods and
Supplementary Material). Blue line: polarization-maintaining (PM) fiber, green
lines: single-mode non- polarization maintaining (SM) fibers. YDFA: ytterbium-
doped fiber amplifer, HWP:  half-wave plate, QWP: quarter-wave plate, PBS:
polarizing beam splitter. {The GaP
crystal is placed above the electron bunch trajectory.} Only the free-space optics  (along the dashed line)
is located near/in the vacuum chamber, the rest (laser source, YDFA and
downstream components) is located in a remote laboratory.} \label{fig:setup_eos} \end{figure*}

In order to increase the signal-to-noise ratio (and thus the dynamic
range), we combined the amplified photonic time-stretch
strategy~\cite{solli2008amplified}, with a balanced detection
technique. The signal is amplified using a home-made ytterbium-doped
fiber amplifier before entering the 2~km fiber. Moreover, at each
electron bunch passage, three laser pulses are sent into the
system (only one of the pulses being modulated by the electron
bunch). The modulated pulse and a reference pulse are subtracted at
the analog level, using a balanced photodetector (see
Figure~\ref{fig:setup_eos}). Furthermore, the second reference pulse
allows a dark reference to be available at the data analysis stage
(see Methods section and supplementary material).

\subsection*{Simultaneous measurement of electron bunch shapes and resulting coherent radiation emission}
A typical single-shot electro-optic signal is represented in
Figure~\ref{fig:exp_results1}a. The data correspond to the longitudinal density profile of the
electron bunch (or more precisely to the electric field in its
vicinity, see Methods and supplementary material). Detailed analysis reveals two components. As expected, a
slowly-varying shape is systematically observed, whose width is of the order
of the electron bunch size. When the electron bunch is ``compressed'' below a
threshold size (technically, by decreasing the momentum compaction factor of
the storage ring), a fast modulation appears on the electro-optic
sampling signal.

\begin{figure*}
\includegraphics[width=18cm]{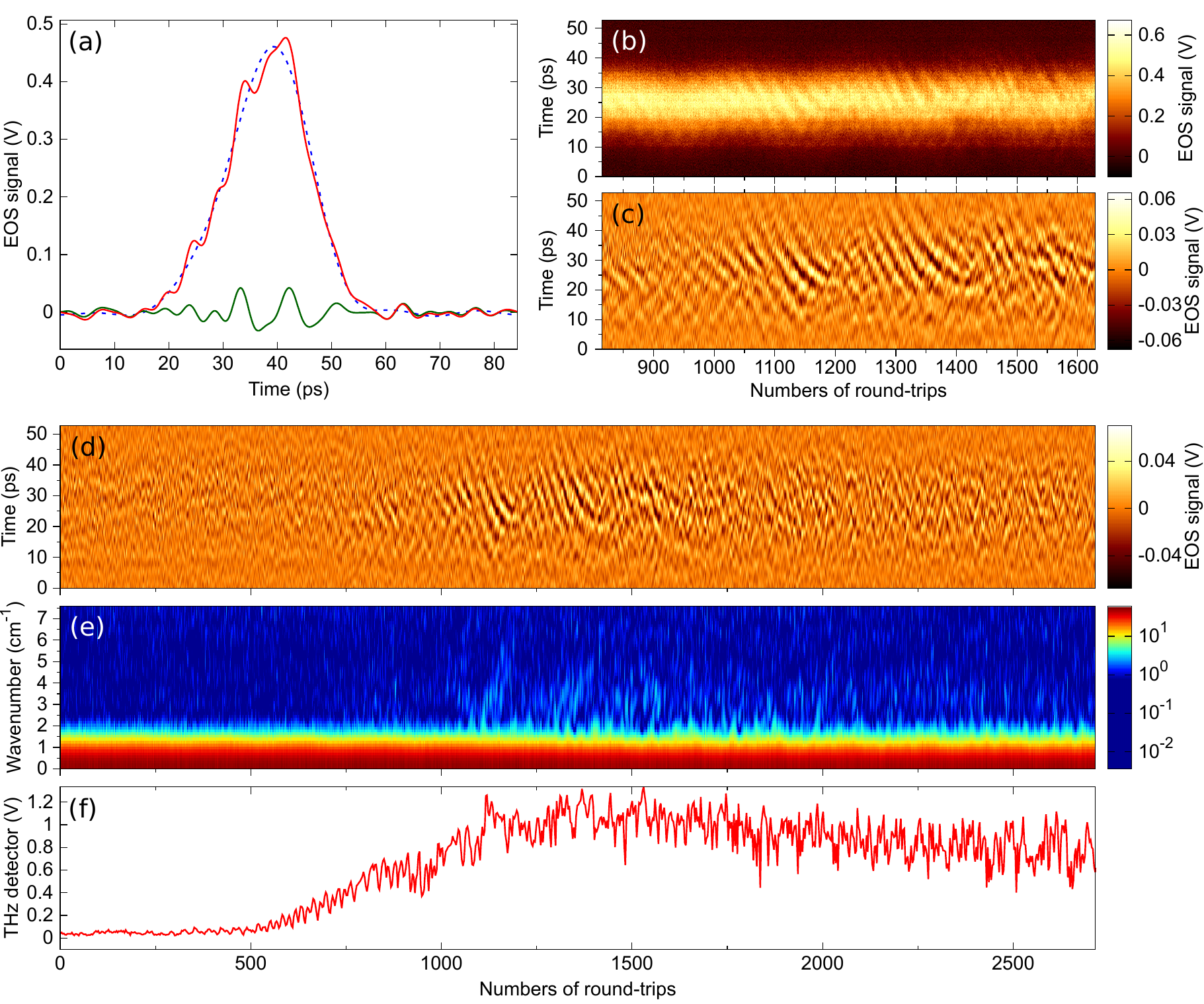}

\caption{{\bf Simultaneous recording of the electron bunch shape at each turn
and associated emission of coherent synchrotron radiation   (CSR).} {\bf (a)}
single-shot recording of an electron bunch shape that   passes near the
detection electro- optic crystal (electric near-field   recorded using time-
stretch electro-optic sampling). Red: electro-optic signal    (over the
0-250~GHz bandwidth). Green: high frequency part between 90 and 250~GHz. Blue:
low frequency part below 80~GHz. {\bf(b,c,d)} Single-shot bunch shapes versus
turns in the storage ring: {\bf(b)} Total electro-optic sampling signal (unfiltered),
{\bf(c)} and {\bf(d)}: high   frequency part (90-250 GHz) revealing the
microbunching structure [{\bf(c)} is a zoom     of {\bf(d)}]. (e) Power
spectrum of each bunch shape versus turn number. {\bf(f)} Emitted
coherent synchrotron radiation recorded   simultaneously at the KARA infrared
beamline using a THz diode   detector (the pulse height is represented at each
turn). Note the correlation between the increase in coherent
synchrotron radiation emission in {\bf(f)} and the spontaneous formation of
microstructures {\bf(d,e)}.} \label{fig:exp_results1} \end{figure*}

In order to conclude non-ambiguously that this structure
corresponds to the microbunching instability, we represented the data
as a function of the revolution number
(Figs.~\ref{fig:exp_results1}b,c,d). {High-pass
  filtered data reveal that the rapidly evolving structure occurs in
  bursts (Figs.~\ref{fig:exp_results1}c), and their space-time
  evolutions (Figs.~\ref{fig:exp_results1}b) present a characteristic
  pattern.} As we will see, this will be a central point for
comparing data with theory.

Since the electro-optic sampling is performed at each turn in
the storage ring, it is possible to examine the correlation of the
spontaneous microstructure formation, with the appearance of coherent
synchrotron radiation. In Figure~\ref{fig:exp_results1}f we have
plotted the data over a long time range, together with the signal
synchronously recorded with a millimeter-wave diode detector placed at
our infrared beamline. We can clearly see the correlation between the
occurence of a burst of CSR, and the growth of the
microstructure. This correlation was systematically observed in the
recorded data.

\section*{Discussion}
{These new data sets can be compared to existing models of electron
bunch dynamics. The physics of the electron bunch evolution involves
essentially three ingredients:} (i) acceleration and energy losses at
each turn, (ii) interaction of each electron with the field created by
the whole electron bunch distribution, (iii) and the relation between
their energy and the revolution time in the storage-ring. The
evolution equation for the distribution of the electrons in phase
space may be written in the form of a Vlasov-Fokker-Planck
equation~\cite{venturini2002.PhysRevLett.89.224802,stupakov2002.PhysRevSTAB.5.054402}:
\begin{equation}
        \frac{\partial f}{\partial\theta} -  p\frac{\partial f}{\partial q} +
        \left[q-I_c E_{\mathit{wf}}(f,q)\right] \frac{\partial f}{\partial p}  = 2\epsilon \frac{\partial
        }{\partial p}\left(pf + \frac{\partial f}{\partial p}\right),
\label{eq:VFP}
\end{equation}
where $f(q,p,\theta)$ is the distribution of the electrons at time $\theta$ in
phase space $(q,p)$. $\theta$ is a continuous and dimensionless variable
associated to the number of turns in the storage ring: $\theta=2\pi f_s t$,
where $t$ is the time (in seconds) and $f_s$ is the synchrotron frequency
(here in the tens of kilohertz range). The longitudinal position $q$ and
relative momentum $p$ are the deviation from the so-called synchronous
electron (with position $z_0$ and energy $E_0$). $q$ and $p$ are expressed in
units of the equilibrium bunch length $\sigma_z$ and energy spread $\sigma_E$
at zero current. $I_cE_{wf}(f,q)$ corresponds to field created by the whole
bunch at the location $q$. We use here only shielded CSR impedance. Details
are given in the Methods section.

{In Figure~\ref{fig:num_results1}, we have represented
  the simulated evolution of the electron bunch shape versus number of
  turns in the storage ring. We can see that this type of
  representation can be used directly for performing severe tests of
  theoretical model versus experimental data. In our case, we can see
  that the model can reproduce part of the spatio-temporal features,
as e.g., structures moving towards the bunch head, and bunch
tail. Evolution versus number of turns also reveals interesting
discrepancies between model predictions and experimental data}. {These types of measurements should allow in due course to refine the models of the wakefiels created by each electron (whose Fourier transform is known as the {\it machine impedance}).}

\begin{figure*}[htbp]
\includegraphics[width=17cm]{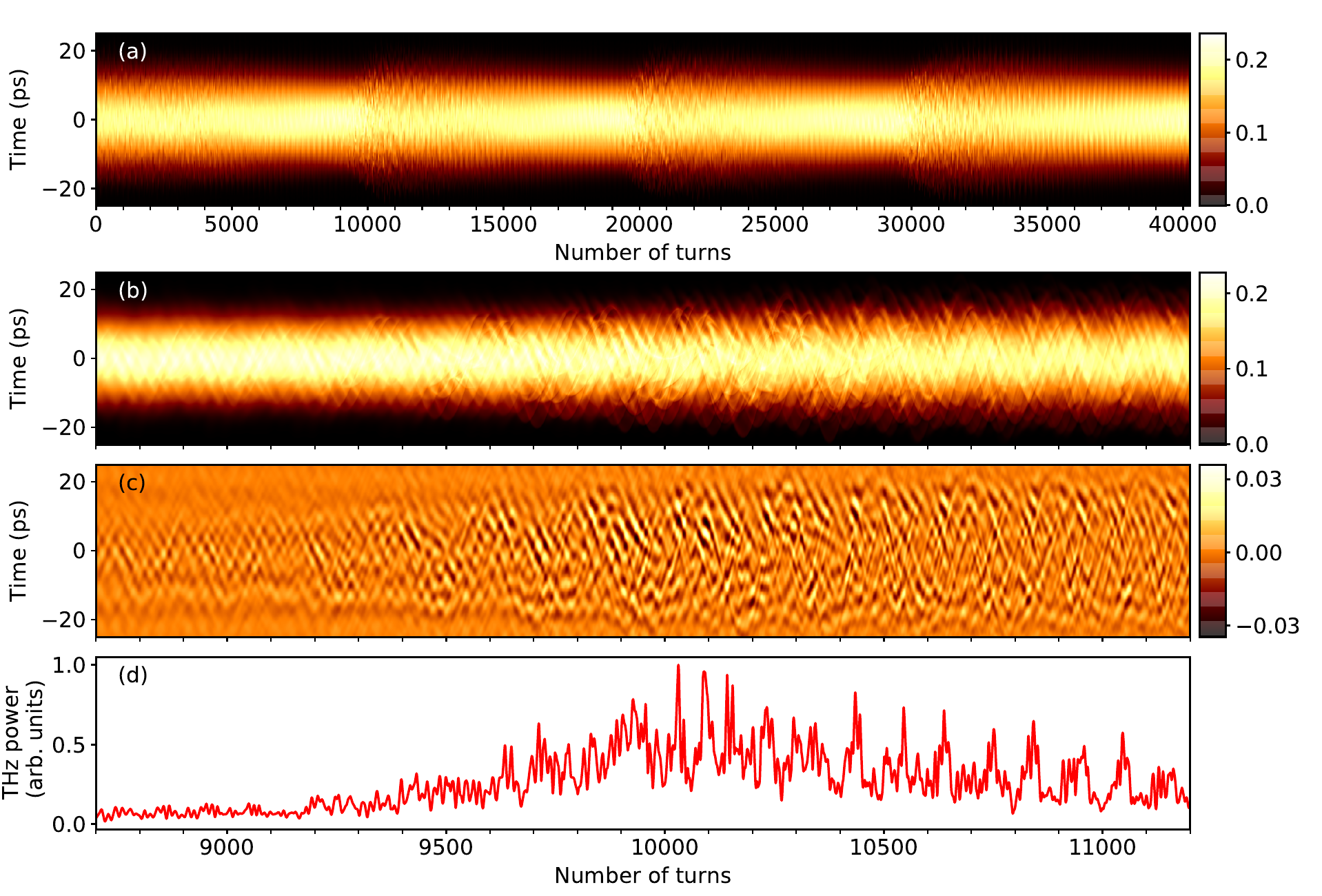} 
\caption{{\bf Numerical
simulation of the electron bunch dynamics.} {\bf (a)} and {\bf (b)} electron bunch shape
at each turn in the storage ring {\bf(b)} is a zoomed view of one of the
bursts of {\bf (a)}. {\bf (c)} filtered data   (in the 90-250~GHz range), revealing the
microstructure evolution. {\bf (d)} Coherent synchrotron radiation emitted by the
microstructure. See Figure~\ref{fig:fig0} for the associated longitudinal density profile and
phase-space at turn~9742, and Supplementary Video~1   for the
corresponding phase space evolution.} 
\label{fig:num_results1} 
\end{figure*}

In conclusion, we present a strategy enabling a simultaneous
measurement of the ``shape'' of electron bunches in a non-destructive
manner at each turn in a storage ring, by monitoring their electric
fields. This new measurement possibility enables to directly observe
the correlation, at each turn, between the charge modulation and the
underlying coherent synchrotron radiation emission, and was predicted
for storage rings more than a decade
ago~\cite{stupakov2002.PhysRevSTAB.5.054402,venturini2002.PhysRevLett.89.224802}. This
type of strategy will enable to start very stringent tests of
theoretical models of relativistic electron bunch dynamics, that were
not possible before. We believe that this direct access to the
microbuching instability and coherent synchrotron radiation properties
may provide an important milestone on the way to master the
instabilities, either for suppressing them, or make them usable as a
stable source of THz radiation.

{In a general way, needs for electron bunch shape diagnostics
  are expected to address challenging questions to the photonic community. An
  important (and related) open question concern the non-destructive
  characterization of electric field oscillations, when the time-scales are in
  the few to tens of microns range. This is an important question for
  studies of microbunching instabilities in lastest generation Free-Electron
  Lasers, and would require to perform single-shot electro-optic sampling of
  mid-infrared pulses. This may represent one of the next milestones in the
  development of photonic systems destined to relativistic electron bunch
  characterization.}

\section*{Methods}
\subsection*{Laser system}
The 1030~nm probe pulses are produced by a mode-locked ytterbium-doped
fiber laser, operating at 62.5~MHz, and synchronized on the RF
reference of the KARA storage ring. An acousto-optic pulse picker
selects 3 pulses per turn in the storage ring. The pulses are
compressed and then amplified in a polarization-maintaining
ytterbium-doped fiber parabolic pulse
amplifier~\cite{thesis_hiller}. The output pulses have a typical
bandwidth of 80~nm FWHM.
\subsection*{Near-field electro-optic sampling setup}
The laser pulses are then transported in a 35~m-long 
polarization maintaining fiber to the electro-optic measurement
system installed in the storage ring. Thanks to a Treacy compressor placed before the fiber, output
pulses can be adjusted in the few tens of ps range. The electro-optic
sampling is performed by an 5~mm-long GaP crystal placed inside the
vacuum chamber{, above the electron beam}. The crystal can be moved towards the electron beam orbit and was placed -- for the data shown -- at a distance of $\approx 4$~mm from the electron bunch. The
quarter-wave plate and half-wave plate (see
Figure~\ref{fig:setup_eos}) are adjusted so that the system is
operated near
extinction~\cite{jiang1999electro,PhysRevSTAB.15.070701}, in order to
obtain high sensitivity. The light exiting the low-power port of the
beam-splitter is injected in a single-mode fiber and transported back
for analysis in the remote photonic analysis station (placed outside
of the storage ring shielding).
\subsection*{Amplified photonic time stretch system}
The modulated chirped laser pulses are first amplified using a
home-made ytterbium-doped fiber preamplifier, and then stretched by
propagation in a 2~km-long single-mode fiber (Corning HI~1060). The
fiber's output is then split using a thin-film 3~dB splitter (see
Figure~\ref{fig:setup_eos}), and the two ports are delayed by exactly
one repetition period of the mode-locked laser (16~ns). Thus a
reference (i.e., unmodulated) laser pulse is subtracted from the laser
pulse which carries the ultrafast modulation using the balanced
photodetector. The balanced photodetector is an InGaAs amplified
photoreceiever (DSC-R412 from Discovery Semiconductors), with a
20~GHz bandwidth. The photoreceiver specifications for gain and noise
are 2800~V/W and 40~pW/$\sqrt{\text{Hz}}$ (both being specified at
1550~nm). The precise delay and relative power levels between the two
photodetector inputs are adjusted using an adjustable delay line and a
variable optical attenuator. Data are recorded using a Lecroy
Labmaster 10~Zi oscilloscope (with an -- overdimensioned -- 30~GHz
bandwidth and 80 Gs/s acquisition rate), and the acquired data are
numerically low-passed filtered at 5~GHz before signal analysis (corresponding to 380~GHz
at the electro-optic crystal location).

{Each recorded pulse is a
replica of the electric field in the near-field of the electron bunch,
which is ``stretched in time'' by a factor $M=75.8$. In other words,
1~ns at oscilloscope input corresponds to 13.2~ps at the
electro-optic crystal. The oscilloscope's 80~gigasamples/s 
acquisition rate corresponds to an effective sampling rate of 6.06~terasamples/s. 
The post-processing filtering to 5~GHz corresponds to an input analog bandwidth limitation of 380~GHz.}

\subsection*{Data processing}
At each turn in the storage ring, three consecutive pulses are emitted
by the laser, and only the last one interacts with the electron bunch
near-field. Thus, at turn $n$, the balanced detector signal contains
four pulses: (i) the raw balanced EOS signal $V^{EOS}_n(t)$, (ii) a
reference balanced signal without EOS modulation $V^{REF}_n(t)$, and
two saturated pulses corresponding to unbalanced pulses (see
supplementary Figures~1-2). The EOS signal represented here corresponds
to $V^{EOS}_n(t)-V^{REF}_n(t)$. The spectra of the EOS signal
(as in Figure~\ref{fig:exp_results1}e) show that a reasonable
signal-to-noise ratio is observed up to 5~GHz bandwidth (i.e., 380~GHz
at input). Hence raw data were first low-pass filtered at 5~GHz,
before data analysis (i.e., a 5~GHz oscilloscope would be sufficient
for the present recording). Then we proceeded to further filtering
for examining the different parts of the spectra. In particular, high frequency
structures of Figs.~\ref{fig:exp_results1}c,d (and the green curve in Fig.~\ref{fig:exp_results1}a) are obtained by
filtering the data in the 1.19~GHz-3.30~GHz band (i.e., 90-250~GHz at
the input). Unfiltered data are represented in Fig.~\ref{fig:exp_results1}b.

{The electro-optic sampling signals (as represented in Fig.~\ref{fig:exp_results1}a-e) hence represent the electric field evolution, multiplied by the laser pulse shape, (see Supplementary material for the signal details).}

\subsection*{Coherent sychrotron radiation analysis}
{The THz pulses are detected at KARA's IR1 infrared
  beamline, using an amplified 140-220~GHz Schotty barrier diode
  detectector (Virginia Diodes Inc. WR5.1ZBD) connected to a 6~GHz
  oscilloscope (Lecroy SDA760ZI-A). Figure~\ref{fig:exp_results1}e
  represents the recorded detector pulse height versus revolution
  number.}

\subsection*{Accelerator parameters}
The results presented in this article are performed in single bunch
operation, at $E=1.287$~GeV energy, for a current $I=1.625$~mA, an
acceleration voltage of $1500$~kV and a momentum compaction factor of
$\alpha=0.724\times 10^{-3}$. The storage ring revolution frequency is
2.716~MHz.

\subsection*{Numerical simulations}
Numerical simulations have been performed using the semi-Lagrangian
scheme~\cite{warnock_scheme_book}, and the shielded CSR wakefield as
in
Refs.~\cite{uvsor_MBI_ybco,uvsor_MBI_ybco,roussel2015.EOS}. Calculations
have been made and cross-checked using two independently developed
codes. One code is a parallel implementation of the Warnock
scheme~\cite{warnock_scheme_book} using MPI (Message
Passing Interface), and the other code is INOVESA which has been developed by the KIT
group~\cite{schonfeldt2017parallelized}. Figure \ref{fig:num_results1}
is provided by the first code, parameters are summarized in the
Supplementary material.

\subsection*{Data availability}
The data that support the findings of this study are available from the
corresponding author upon reasonable request.

\section*{Acknowledgments}
{This work has been supported by the German Federal
  Ministry of Education and Research (contract no. 05K16VKA) and by
  the Initiative and Networking Fund of the Helmholtz Association
  (contract no. VH-NG-320). On the PhLAM side, the work has been
  supported by the Ministry of Higher Education and Research, Nord-Pas
  de Calais Regional Council and Eu\-ropean Regional Development Fund
  (ERDF) through the Contrat de Projets \'Etat-R\'egion (CPER
  photonics for society), and the LABEX CEMPI project (ANR-11-
  LABX-0007). This work was performed using HPC resources from GENCI-IDRIS
   (Grants i2015057057, i2016057057, A0040507057).}

\section*{Author contributions}
{{Main project management has been peformed by ASM.} 
The association of the
EOS system with photonic time-stretch has been developed by NH, EBlo,
SF, EBru, MJN, GN, PS, MS, JLS, SW, and ASM on the EOS side. The amplified
time-stretch readout has been developed by CE, ER, CS,
SB. Time-stretch data analysis has been performed by CS and CE.
Measurement of the THz radiation has been performed by JLS and MB. EBlo
and MS prepared the KARA fill with low-alpha optics. Numerical
simulations have been performed by EBur, PS, CE, SB, and simulation code
development by ER, CE, SB (MPI implementation of the Warnock scheme), and PS
(INOVESA code). All authors participated in the manuscript redaction.

\section*{Competing Interests}
 The authors have no competing interests as defined by Nature Research, or other interests that might be perceived to influence the results and/or discussion reported in this paper.


\newpage

\begin{thebibliography}{10}
\expandafter\ifx\csname url\endcsname\relax
  \def\url#1{\texttt{#1}}\fi
\expandafter\ifx\csname urlprefix\endcsname\relax\def\urlprefix{URL }\fi
\providecommand{\bibinfo}[2]{#2}
\providecommand{\eprint}[2][]{\url{#2}}

\bibitem{hight1997ar}
\bibinfo{author}{Hight-Walker, A.}, \bibinfo{author}{Arp, U.},
  \bibinfo{author}{Fraser, G.}, \bibinfo{author}{Lucatorto, T.} \&
  \bibinfo{author}{Wen, J.}
\newblock New infrared beamline at the NIST SURF II storage ring.
\newblock \emph{\bibinfo{journal}{Proc. SPIE Int. Soc. Opt. Eng.}}
  \textbf{\bibinfo{volume}{3153}}, \bibinfo{pages}{40} (\bibinfo{year}{1997}).

\bibitem{andersson2000coherent}
\bibinfo{author}{Andersson, A.}, \bibinfo{author}{Johnson, M.~S.} \&
  \bibinfo{author}{Nelander, B.}
\newblock Coherent synchrotron radiation in the far-infrared from a 1~mm
  electron bunch.
\newblock \emph{\bibinfo{journal}{Optical Engineering}}
  \textbf{\bibinfo{volume}{39}}, \bibinfo{pages}{3099--3106}
  (\bibinfo{year}{2000}).

\bibitem{byrd2002.PhysRevLett.89.224801}
\bibinfo{author}{Byrd, J.~M.} \emph{et~al.}
\newblock Observation of Broadband Self-Amplified Spontaneous Coherent
  Terahertz Synchrotron Radiation in a Storage Ring.
\newblock \emph{\bibinfo{journal}{Phys. Rev. Lett.}}
  \textbf{\bibinfo{volume}{89}}, \bibinfo{pages}{224801}
  (\bibinfo{year}{2002}).

\bibitem{abo2002.PhysRevLett.88.254801}
\bibinfo{author}{Abo-Bakr, M.}, \bibinfo{author}{Feikes, J.},
  \bibinfo{author}{Holldack, K.}, \bibinfo{author}{W\"ustefeld, G.} \&
  \bibinfo{author}{H\"ubers, H.-W.}
\newblock Steady-State Far-Infrared Coherent Synchrotron Radiation detected at
  {BESSY {II}}.
\newblock \emph{\bibinfo{journal}{Phys. Rev. Lett.}}
  \textbf{\bibinfo{volume}{88}}, \bibinfo{pages}{254801}
  (\bibinfo{year}{2002}).

\bibitem{takashima2005observation}
\bibinfo{author}{Takashima, Y.} \emph{et~al.}
\newblock Observation of intense bursts of terahertz synchrotron radiation at
  uvsor-ii.
\newblock \emph{\bibinfo{journal}{Japanese journal of applied physics}}
  \textbf{\bibinfo{volume}{44}}, \bibinfo{pages}{L1131} (\bibinfo{year}{2005}).

\bibitem{katoh2007coherent}
\bibinfo{author}{Katoh, M.} \emph{et~al.}
\newblock Coherent Terahertz Radiation at UVSOR-II.
\newblock In \emph{\bibinfo{booktitle}{AIP Conference Proceedings}}, vol.
  \bibinfo{volume}{879}, \bibinfo{pages}{71--74} (\bibinfo{organization}{AIP},
  \bibinfo{year}{2007}).

\bibitem{Karantzoulis2010300}
\bibinfo{author}{Karantzoulis, E.}, \bibinfo{author}{Penco, G.},
  \bibinfo{author}{Perucchi, A.} \& \bibinfo{author}{Lupi, S.}
\newblock Characterization of coherent {THz} radiation bursting regime at
  {ELETTRA}.
\newblock \emph{\bibinfo{journal}{Infrared Physics and Technology}}
  \textbf{\bibinfo{volume}{53}}, \bibinfo{pages}{300} (\bibinfo{year}{2010}).

\bibitem{muller2011metrology}
\bibinfo{author}{M{\"u}ller, R.} \emph{et~al.}
\newblock The Metrology Light Source of PTB--a source for THz radiation.
\newblock \emph{\bibinfo{journal}{Journal of Infrared, Millimeter, and
  Terahertz Waves}} \textbf{\bibinfo{volume}{32}}, \bibinfo{pages}{742--753}
  (\bibinfo{year}{2011}).

\bibitem{Feikes2011metrology}
\bibinfo{author}{Feikes, J.} \emph{et~al.}
\newblock Metrology Light Source: The first electron storage ring optimized for
  generating coherent THz radiation.
\newblock \emph{\bibinfo{journal}{Physical Review Special Topics-Accelerators
  and Beams}} \textbf{\bibinfo{volume}{14}}, \bibinfo{pages}{030705}
  (\bibinfo{year}{2011}).

\bibitem{cinque2011far}
\bibinfo{author}{Cinque, G.}, \bibinfo{author}{Frogley, M.~D.} \&
  \bibinfo{author}{Bartolini, R.}
\newblock Far-IR/THz spectral characterization of the coherent synchrotron
  radiation emission at diamond IR beamline B22.
\newblock \emph{\bibinfo{journal}{Rendiconti Lincei}}
  \textbf{\bibinfo{volume}{22}}, \bibinfo{pages}{33--47}
  (\bibinfo{year}{2011}).

\bibitem{shields2012.1742-6596-357-1-012037}
\bibinfo{author}{Shields, W.} \emph{et~al.}
\newblock Microbunch Instability Observations from a {THz} Detector at Diamond
  Light Source.
\newblock \emph{\bibinfo{journal}{Journal of Physics: Conference Series}}
  \textbf{\bibinfo{volume}{357}}, \bibinfo{pages}{012037}
  (\bibinfo{year}{2012}).

\bibitem{evain2012.0295-5075-98-4-40006}
\bibinfo{author}{Evain, C.} \emph{et~al.}
\newblock Spatio-temporal dynamics of relativistic electron bunches during the
  micro-bunching instability in storage rings.
\newblock \emph{\bibinfo{journal}{EPL}} \textbf{\bibinfo{volume}{98}},
  \bibinfo{pages}{40006} (\bibinfo{year}{2012}).

\bibitem{mullerexperimentalapects}
\bibinfo{author}{M\"uller, A.~S.} \emph{et~al.}
\newblock Experimental Aspects of CSR in the ANKA Storage Ring.
\newblock \emph{\bibinfo{journal}{ICFA Beam Dynamics Newsletter}}
  \textbf{\bibinfo{volume}{57}}, \bibinfo{pages}{154} (\bibinfo{year}{2012}).

\bibitem{barros2015characteristics}
\bibinfo{author}{Barros, J.} \emph{et~al.}
\newblock Characteristics and development of the coherent synchrotron radiation
  sources for THz spectroscopy.
\newblock \emph{\bibinfo{journal}{Journal of Molecular Spectroscopy}}
  \textbf{\bibinfo{volume}{315}}, \bibinfo{pages}{3--9} (\bibinfo{year}{2015}).

\bibitem{roy.RSI.84.033102.2013}
\bibinfo{author}{Barros, J.} \emph{et~al.}
\newblock Coherent synchrotron radiation for broadband terahertz spectroscopy.
\newblock \emph{\bibinfo{journal}{Review of Scientific Instruments}}
  \textbf{\bibinfo{volume}{84}}, \bibinfo{pages}{033102}
  (\bibinfo{year}{2013}).

\bibitem{PhysRevLett.114.204801}
\bibinfo{author}{Billinghurst, B.~E.} \emph{et~al.}
\newblock Observation of Wakefields and Resonances in Coherent Synchrotron
  Radiation.
\newblock \emph{\bibinfo{journal}{Phys. Rev. Lett.}}
  \textbf{\bibinfo{volume}{114}}, \bibinfo{pages}{204801}
  (\bibinfo{year}{2015}).
\newblock
  \urlprefix\url{http://link.aps.org/doi/10.1103/PhysRevLett.114.204801}.

\bibitem{steinmann2018continuous}
\bibinfo{author}{Steinmann, J.~L.} \emph{et~al.}
\newblock Continuous bunch-by-bunch spectroscopic investigation of the
  microbunching instability.
\newblock \emph{\bibinfo{journal}{Physical Review Accelerators and Beams}}
  \textbf{\bibinfo{volume}{21}}, \bibinfo{pages}{110705}
  (\bibinfo{year}{2018}).

\bibitem{venturini2002.PhysRevLett.89.224802}
\bibinfo{author}{Venturini, M.} \& \bibinfo{author}{Warnock, R.}
\newblock Bursts of Coherent Synchrotron Radiation in Electron Storage Rings: A
  Dynamical Model.
\newblock \emph{\bibinfo{journal}{Phys. Rev. Lett.}}
  \textbf{\bibinfo{volume}{89}}, \bibinfo{pages}{224802}
  (\bibinfo{year}{2002}).

\bibitem{stupakov2002.PhysRevSTAB.5.054402}
\bibinfo{author}{Stupakov, G.} \& \bibinfo{author}{Heifets, S.}
\newblock Beam instability and microbunching due to coherent synchrotron
  radiation.
\newblock \emph{\bibinfo{journal}{Phys. Rev. ST Accel. Beams}}
  \textbf{\bibinfo{volume}{5}}, \bibinfo{pages}{054402} (\bibinfo{year}{2002}).

\bibitem{cross1993.RevModPhys.65.851}
\bibinfo{author}{Cross, M.~C.} \& \bibinfo{author}{Hohenberg, P.~C.}
\newblock Pattern formation outside of equilibrium.
\newblock \emph{\bibinfo{journal}{Rev. Mod. Phys.}}
  \textbf{\bibinfo{volume}{65}}, \bibinfo{pages}{851--1112}
  (\bibinfo{year}{1993}).

\bibitem{tai1986observation}
\bibinfo{author}{Tai, K.}, \bibinfo{author}{Hasegawa, A.} \&
  \bibinfo{author}{Tomita, A.}
\newblock Observation of modulational instability in optical fibers.
\newblock \emph{\bibinfo{journal}{Physical review letters}}
  \textbf{\bibinfo{volume}{56}}, \bibinfo{pages}{135} (\bibinfo{year}{1986}).

\bibitem{charru2013sand}
\bibinfo{author}{Charru, F.}, \bibinfo{author}{Andreotti, B.} \&
  \bibinfo{author}{Claudin, P.}
\newblock Sand ripples and dunes.
\newblock \emph{\bibinfo{journal}{Annual Review of Fluid Mechanics}}
  \textbf{\bibinfo{volume}{45}}, \bibinfo{pages}{469--493}
  (\bibinfo{year}{2013}).

\bibitem{helbing2001traffic}
\bibinfo{author}{Helbing, D.}
\newblock Traffic and related self-driven many-particle systems.
\newblock \emph{\bibinfo{journal}{Reviews of modern physics}}
  \textbf{\bibinfo{volume}{73}}, \bibinfo{pages}{1067} (\bibinfo{year}{2001}).

\bibitem{EOS_first_spectral_encoding}
\bibinfo{author}{Jiang, Z.} \& \bibinfo{author}{Zhang, X.-C.}
\newblock Electro-optic measurement of {THz} field pulses with a chirped
  optical beam.
\newblock \emph{\bibinfo{journal}{Appl. Phys. Letters}}
  \textbf{\bibinfo{volume}{72}}, \bibinfo{pages}{1945} (\bibinfo{year}{1998}).

\bibitem{PhysRevLett.88.124801}
\bibinfo{author}{Wilke, I.} \emph{et~al.}
\newblock Single-Shot Electron-Beam Bunch Length Measurements.
\newblock \emph{\bibinfo{journal}{Phys. Rev. Lett.}}
  \textbf{\bibinfo{volume}{88}}, \bibinfo{pages}{124801}
  (\bibinfo{year}{2002}).

\bibitem{casalbuoni2008numerical}
\bibinfo{author}{Casalbuoni, S.} \emph{et~al.}
\newblock Numerical studies on the electro-optic detection of femtosecond
  electron bunches.
\newblock \emph{\bibinfo{journal}{Physical Review Special Topics-Accelerators
  and Beams}} \textbf{\bibinfo{volume}{11}}, \bibinfo{pages}{072802}
  (\bibinfo{year}{2008}).

\bibitem{steffen2009electro}
\bibinfo{author}{Steffen, B.} \emph{et~al.}
\newblock Electro-optic time profile monitors for femtosecond electron bunches
  at the soft x-ray free-electron laser FLASH.
\newblock \emph{\bibinfo{journal}{Physical Review Special Topics-Accelerators
  and Beams}} \textbf{\bibinfo{volume}{12}}, \bibinfo{pages}{032802}
  (\bibinfo{year}{2009}).

\bibitem{schmidhammer2009single}
\bibinfo{author}{Schmidhammer, U.}, \bibinfo{author}{De~Waele, V.},
  \bibinfo{author}{Marques, J.-R.}, \bibinfo{author}{Bourgeois, N.} \&
  \bibinfo{author}{Mostafavi, M.}
\newblock Single shot linear detection of 0.01--10 THz electromagnetic fields.
\newblock \emph{\bibinfo{journal}{Applied Physics B}}
  \textbf{\bibinfo{volume}{94}}, \bibinfo{pages}{95} (\bibinfo{year}{2009}).

\bibitem{PhysRevSTAB.15.070701}
\bibinfo{author}{M\"uller, F.} \emph{et~al.}
\newblock Electro-optical measurement of sub-ps structures in low charge
  electron bunches.
\newblock \emph{\bibinfo{journal}{Phys. Rev. ST Accel. Beams}}
  \textbf{\bibinfo{volume}{15}}, \bibinfo{pages}{070701}
  (\bibinfo{year}{2012}).

\bibitem{szwaj2016high}
\bibinfo{author}{Szwaj, C.} \emph{et~al.}
\newblock High sensitivity photonic time-stretch electro-optic sampling of
  terahertz pulses.
\newblock \emph{\bibinfo{journal}{Review of Scientific Instruments}}
  \textbf{\bibinfo{volume}{87}}, \bibinfo{pages}{103111}
  (\bibinfo{year}{2016}).

\bibitem{tamura1996pulse}
\bibinfo{author}{Tamura, K.} \& \bibinfo{author}{Nakazawa, M.}
\newblock Pulse compression by nonlinear pulse evolution with reduced optical
  wave breaking in erbium-doped fiber amplifiers.
\newblock \emph{\bibinfo{journal}{Optics letters}}
  \textbf{\bibinfo{volume}{21}}, \bibinfo{pages}{68--70}
  (\bibinfo{year}{1996}).

\bibitem{fermann2000self}
\bibinfo{author}{Fermann, M.}, \bibinfo{author}{Kruglov, V.},
  \bibinfo{author}{Thomsen, B.}, \bibinfo{author}{Dudley, J.} \&
  \bibinfo{author}{Harvey, J.}
\newblock Self-similar propagation and amplification of parabolic pulses in
  optical fibers.
\newblock \emph{\bibinfo{journal}{Physical Review Letters}}
  \textbf{\bibinfo{volume}{84}}, \bibinfo{pages}{6010} (\bibinfo{year}{2000}).

\bibitem{feurer2009ytterbium}
\bibinfo{author}{Muller, F.} \emph{et~al.}
\newblock Ytterbium fiber laser for electro-optical pulse length measurements
  at the SwissFEL.
\newblock \emph{\bibinfo{journal}{Proceedings of DIPAC09, TUPD31}}
  (\bibinfo{year}{2009}).

\bibitem{jalali2015tailoring}
\bibinfo{author}{Jalali, B.} \& \bibinfo{author}{Mahjoubfar, A.}
\newblock Tailoring wideband signals with a photonic hardware accelerator.
\newblock \emph{\bibinfo{journal}{Proceedings of the IEEE}}
  \textbf{\bibinfo{volume}{103}}, \bibinfo{pages}{1071--1086}
  (\bibinfo{year}{2015}).

\bibitem{mahjoubfar2017time}
\bibinfo{author}{Mahjoubfar, A.} \emph{et~al.}
\newblock Time stretch and its applications.
\newblock \emph{\bibinfo{journal}{Nature Photonics}}
  \textbf{\bibinfo{volume}{11}}, \bibinfo{pages}{341--351}
  (\bibinfo{year}{2017}).

\bibitem{time_stretch_first_bhushan1998time}
\bibinfo{author}{Bhushan, A.}, \bibinfo{author}{Coppinger, F.} \&
  \bibinfo{author}{Jalali, B.}
\newblock Time-stretched analogue-to-digital conversion.
\newblock \emph{\bibinfo{journal}{Electronics Letters}}
  \textbf{\bibinfo{volume}{34}}, \bibinfo{pages}{1081--1083}
  (\bibinfo{year}{1998}).

\bibitem{roussel2015.EOS}
\bibinfo{author}{Roussel, E.} \emph{et~al.}
\newblock Observing microscopic structures of a relativistic object using a
  time-stretch strategy.
\newblock \emph{\bibinfo{journal}{Scientific Reports}}
  \textbf{\bibinfo{volume}{5}} (\bibinfo{year}{2015}).

\bibitem{kobayashi2016high}
\bibinfo{author}{Kobayashi, M.} \emph{et~al.}
\newblock High-Acquisition-Rate Single-Shot Pump-Probe Measurements Using
  Time-Stretching Method.
\newblock \emph{\bibinfo{journal}{Scientific reports}}
  \textbf{\bibinfo{volume}{6}}, \bibinfo{pages}{37614} (\bibinfo{year}{2016}).

\bibitem{evain2017direct}
\bibinfo{author}{Evain, C.} \emph{et~al.}
\newblock Direct observation of spatiotemporal dynamics of short electron
  bunches in storage rings.
\newblock \emph{\bibinfo{journal}{Physical Review Letters}}
  \textbf{\bibinfo{volume}{118}}, \bibinfo{pages}{054801}
  (\bibinfo{year}{2017}).

\bibitem{kramermicrowave}
\bibinfo{author}{Kramer, S.}, \bibinfo{author}{Carr, G.} \&
  \bibinfo{author}{Podobedov, B.}
\newblock Microwave Emission Measurements from the Electron Beam in the {VUV}
  Ring.
\newblock \bibinfo{note}{NSLS activity report (2001)}.

\bibitem{uvsor_MBI_ybco}
\bibinfo{author}{Roussel, E.} \emph{et~al.}
\newblock Microbunching Instability in Relativistic Electron Bunches: Direct
  Observations of the Microstructures Using Ultrafast {YBCO} Detectors.
\newblock \emph{\bibinfo{journal}{Phys. Rev. Lett.}}
  \textbf{\bibinfo{volume}{113}}, \bibinfo{pages}{094801}
  (\bibinfo{year}{2014}).

\bibitem{ANKA.EOS.IPAC13}
\bibinfo{author}{Hiller, N.} \emph{et~al.}
\newblock A.
\newblock In \emph{\bibinfo{booktitle}{Proceedings of the 2013 Particle
  Accelerator Conference, Shanghai, China}}, \bibinfo{pages}{MOPME014}
  (\bibinfo{year}{2013}).

\bibitem{thesis_hiller}
\bibinfo{author}{Hiller, N.}
\newblock \bibinfo{note}{“Electro-Optical Bunch Length Measurements at the
  ANKA Storage Ring”, PhD thesis, KIT, 2013.}

\bibitem{rota2018kalypso}
\bibinfo{author}{Rota, L.} \emph{et~al.}
\newblock {KALYPSO}: Linear array detector for high-repetition rate and
  real-time beam diagnostics.
\newblock \emph{\bibinfo{journal}{To appear in Nuclear Instruments and Methods
  in Physics Research Section A}}  (\bibinfo{year}{2018}).

\bibitem{kalypso_microstructures_first}
\bibinfo{author}{Rota, L.} \emph{et~al.}
\newblock {KALYPSO}: A {Mfps} linear array detector for visible to near
  radiation.
\newblock \bibinfo{note}{Proceedings of IBIC, WEPG46 (2016)}.

\bibitem{kehrer2018synchronous}
\bibinfo{author}{Kehrer, B.} \emph{et~al.}
\newblock Synchronous detection of longitudinal and transverse bunch signals at
  a storage ring.
\newblock \emph{\bibinfo{journal}{Physical Review Accelerators and Beams}}
  \textbf{\bibinfo{volume}{21}}, \bibinfo{pages}{102803}
  (\bibinfo{year}{2018}).

\bibitem{funkner2018high}
\bibinfo{author}{Funkner, S.} \emph{et~al.}
\newblock High throughput data streaming of individual longitudinal electron
  bunch profiles in a storage ring with single-shot electro-optical sampling.
\newblock \emph{\bibinfo{journal}{arXiv preprint arXiv:1809.07530}}
  (\bibinfo{year}{2018}).

\bibitem{solli2008amplified}
\bibinfo{author}{Solli, D.}, \bibinfo{author}{Chou, J.} \&
  \bibinfo{author}{Jalali, B.}
\newblock Amplified wavelength--time transformation for real-time spectroscopy.
\newblock \emph{\bibinfo{journal}{Nature Photonics}}
  \textbf{\bibinfo{volume}{2}}, \bibinfo{pages}{48--51} (\bibinfo{year}{2008}).

\bibitem{jiang1999electro}
\bibinfo{author}{Jiang, Z.}, \bibinfo{author}{Sun, F.}, \bibinfo{author}{Chen,
  Q.} \& \bibinfo{author}{Zhang, X.-C.}
\newblock Electro-optic sampling near zero optical transmission point.
\newblock \emph{\bibinfo{journal}{Applied physics letters}}
  \textbf{\bibinfo{volume}{74}}, \bibinfo{pages}{1191--1193}
  (\bibinfo{year}{1999}).

\bibitem{warnock_scheme_book}
\bibinfo{author}{Warnock, R.} \& \bibinfo{author}{Ellison, J.}
\newblock In \emph{\bibinfo{booktitle}{Proceedings of the 2nd ICFA advanced
  accelerator workshop, The physics of high brightness beams, World Scientific,
  Singapore}}, \bibinfo{pages}{322} (\bibinfo{year}{2000}).

\bibitem{schonfeldt2017parallelized}
\bibinfo{author}{Sch{\"o}nfeldt, P.}, \bibinfo{author}{Brosi, M.},
  \bibinfo{author}{Schwarz, M.}, \bibinfo{author}{Steinmann, J.~L.} \&
  \bibinfo{author}{M{\"u}ller, A.-S.}
\newblock Parallelized Vlasov-Fokker-Planck solver for desktop personal
  computers.
\newblock \emph{\bibinfo{journal}{Physical Review Accelerators and Beams}}
  \textbf{\bibinfo{volume}{20}}, \bibinfo{pages}{030704}
  (\bibinfo{year}{2017}).

\end{thebibliography}
\newpage
\end{document}